\documentclass[12pt,prl,aps,epsfig,psfig]{revtex4}
\usepackage{epsfig,epsf}
\begin{document}                                                              
\newcommand{\beq}{\begin{equation}}
\newcommand{\eeq}{\end{equation}}

\newcommand{\be}{\begin{eqnarray}}
\newcommand{\ee}{\end{eqnarray}}
\newcommand{\dd}{\mathrm{d\,}}
\def\eq#1{{Eq.~(\ref{#1})}}
\def\fig#1{{Fig.~\ref{#1}}}
\newcommand{\as}{\alpha_S}
\newcommand{\bra}[1]{\langle #1 |}
\newcommand{\ket}[1]{|#1\rangle}   
\newcommand{\bracket}[2]{\langle #1|#2\rangle}
\newcommand{\intp}[1]{\int \frac{d^4 #1}{(2\pi)^4}}
\newcommand{\mn}{{\mu\nu}}
\newcommand{\ab}{{\alpha\beta}}
\newcommand{\tr}{{\rm tr}}
\newcommand{\Tr}{{\rm Tr}}
\newcommand{\T} {\mbox{T}}
\newcommand{\braket}[2]{\langle #1|#2\rangle}

\title{Jet Azimuthal Correlations and Parton Saturation \\ in the Color
Glass Condensate}

\author{Dmitri Kharzeev$^a$,  Eugene Levin$^{b}$ and Larry McLerran$^a$}

\bigskip
 
\affiliation{
a) Department of Physics, Brookhaven National Laboratory,\\
Upton, New York 11973-5000, USA\\
b) HEP Department, School of Physics,\\
Raymond and Beverly Sackler Faculty of Exact Science,\\
Tel Aviv University, Tel Aviv 69978, Israel\\
}

\date{\today}
\pacs{}

\begin{abstract}

We consider the influence of parton saturation in the Color Glass
Condensate on the back-to-back azimuthal correlations of high $p_T$
hadrons in $pA$ (or $dA$) collisions. When both near--side and away--side
hadrons are detected at mid-rapidity at RHIC energy, the effects of parton
saturation are constrained to transverse momenta below the saturation
scale $p_T \leq Q_s$; in this case the back-to-back correlations do not
disappear but exhibit broadening. However when near-side and away-side
hadrons are separated by several units of rapidity, quantum evolution
effects lead to the depletion of back-to-back correlations as a function
of rapidity interval between the detected hadrons (at fixed $p_T$). This
applies to both $pp$ and $pA$ (or $dA$) collisions; however, due to the
initial conditions provided by the Color Glass Condensate, the depletion
of the back-to-back correlations is significantly stronger in the $pA$
case.  An experimental study of this effect would thus help to clarify the
origin of the high $p_T$ hadron suppression at forward rapidities observed
recently at RHIC.

\end{abstract}
\maketitle


Recently, a strong suppression of the high $p_T$ hadron yields has been
observed at forward rapidities at RHIC \cite{RHICres1,RHICres2,RHICres3,RHICres4}. 
 Since this effect
has been predicted \cite{KLM,KKT,Alb,BKW} as a signature of quantum
evolution in the Color Glass Condensate \cite{GLR,hdQCD,MV,IF}, the observations have
excited considerable interest. In this paper we consider an observable
which allows to test further the origin of the observed effect -- the
azimuthal back-to-back correlations of high $p_T$ hadrons. To do this, we
extend the KLMN approach \cite{KN,KL,KLN,KLM,KLNDA}, that has been
developed to describe the experimental data on hadron multiplicities and
the inclusive high $p_T$ yields, to the azimuthal correlations
\cite{RHICAZ}.

\begin{figure}[h] 
\begin{minipage}{10cm} 
\begin{center} 
\epsfxsize=9cm
\leavevmode 
\hbox{ \epsffile{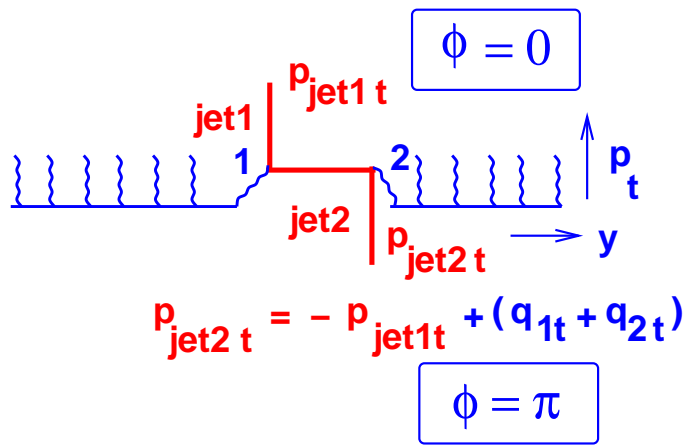}} 
\end{center} 
\end{minipage}
\begin{minipage}{6cm} 
\caption{Back-to-back correlations of produced jets
in the perturbative phase based on the QCD factorization.}
\label{cor1}
\end{minipage}
\end{figure}

The azimuthal correlations provide a powerful method for the diagnostics
of a partonic system (for a recent treatment of azimuthal correlations in nuclear 
collisions, see \cite{KT}).  As we will show in this paper, the measurements of
the strength of back-to-back correlations allow one to tell whether the
partonic system under study has reached the density needed for the
formation of Color Glass Condensate (CGC), or whether it is still in the
perturbative QCD (pQCD) phase. Indeed, in leading order pQCD a typical
hard scattering process at high energy is composed of a gluon jet with a
large transverse momentum ($p_{1,t}$) balanced in the opposite direction
by another gluon jet with transverse momentum ($p_{2,t}$) which is also
large and almost compensates the value of $p_{1,t}$, namely,
$\vec{p}_{2,t} -\vec{p}_{1,t}\,\,=\,\,\vec{q}_1 + \vec{q}_2
\,\ll\,\,|p_{1,t}|$ ( see \fig{cor1}).  However, in the CGC phase of QCD
the phenomenon of saturation implies a different structure of the event: a
jet with large transverse momentum can be compensated by the production of
several gluons with the average transverse momenta which are about equal
to the saturation scale ($Q_s$)  ($p_{2,t} \,\approx\,Q_s$).

In the transitional region dominated by quantum evolution (``Color Quantum
Fluid'' phase or region of extended scaling),  
the dynamics is driven by the interplay of these two
mechanisms, which we are now going to discuss in more detail.
  
\begin{figure}[h]
\begin{center}
\epsfxsize=17cm
\leavevmode
\hbox{ \epsffile{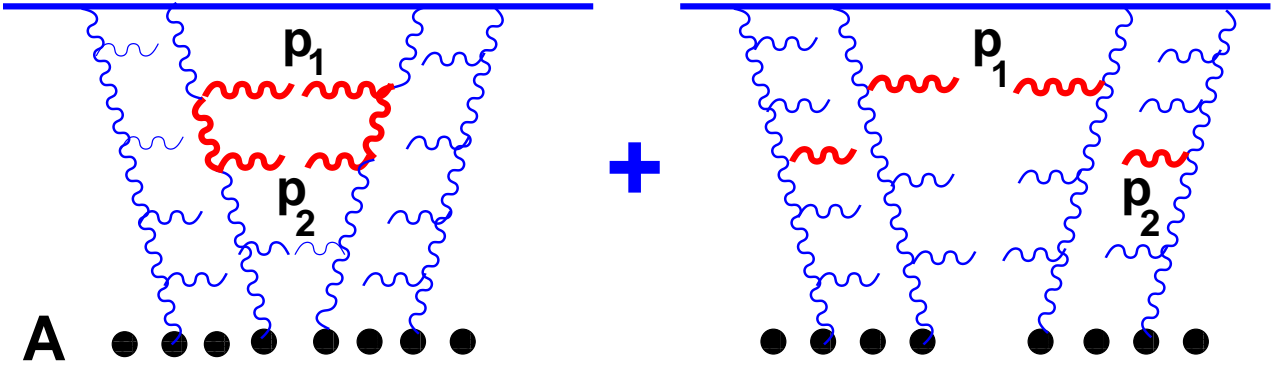}}
\end{center}
\caption{Two mechanisms of double inclusive production. }
 \label{jetcor}
\end{figure}

The first one is the production of two gluon jets from one parton shower
(see the first diagram of \fig{jetcor}) while the second mechanism is the
production of two jets from different parton showers (see the second
diagram of \fig{jetcor}).

\begin{figure}[h] 
\begin{tabular}{c c} 
\epsfig{file=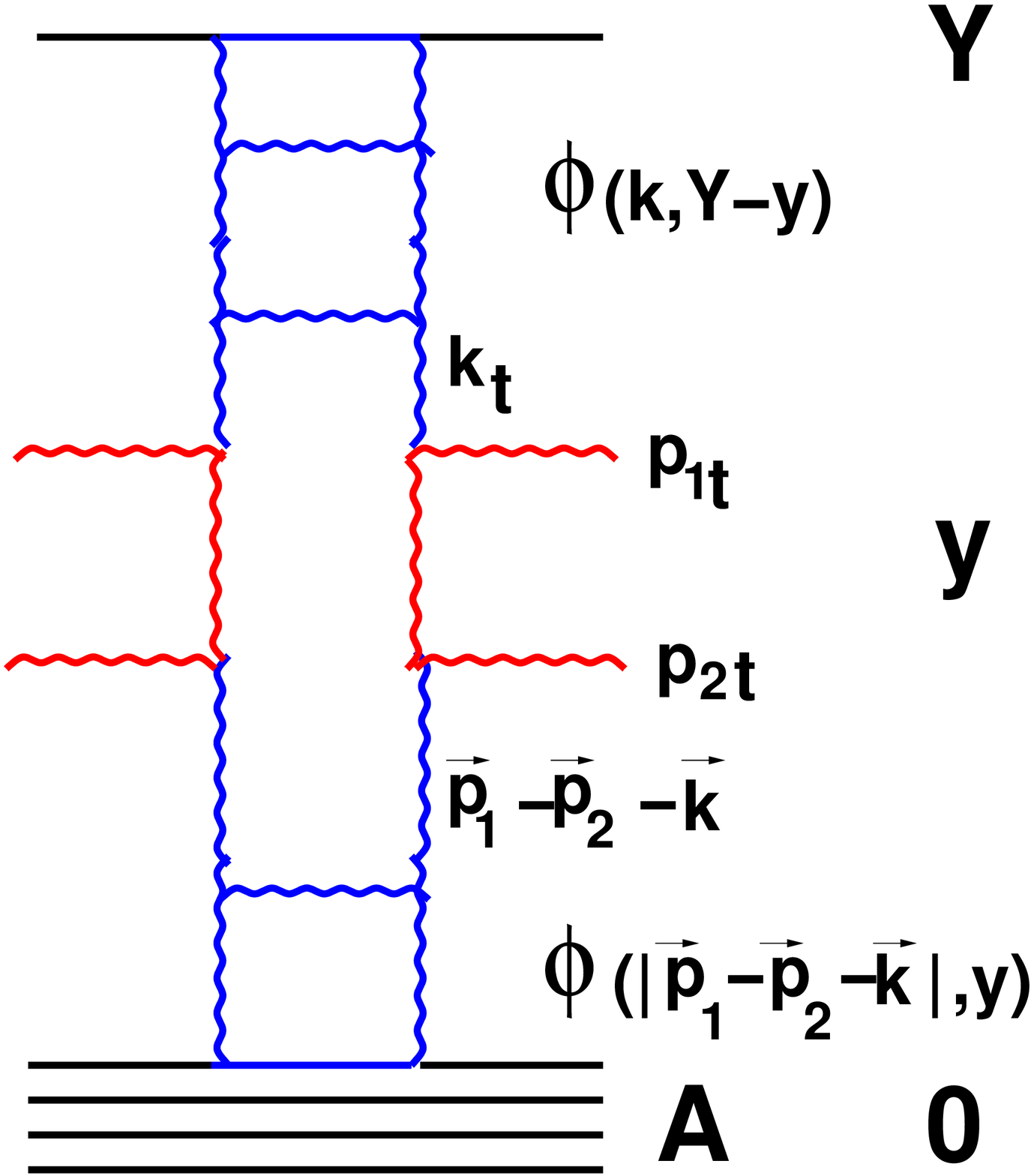,width=8.5cm,height=8cm} &
\epsfig{file=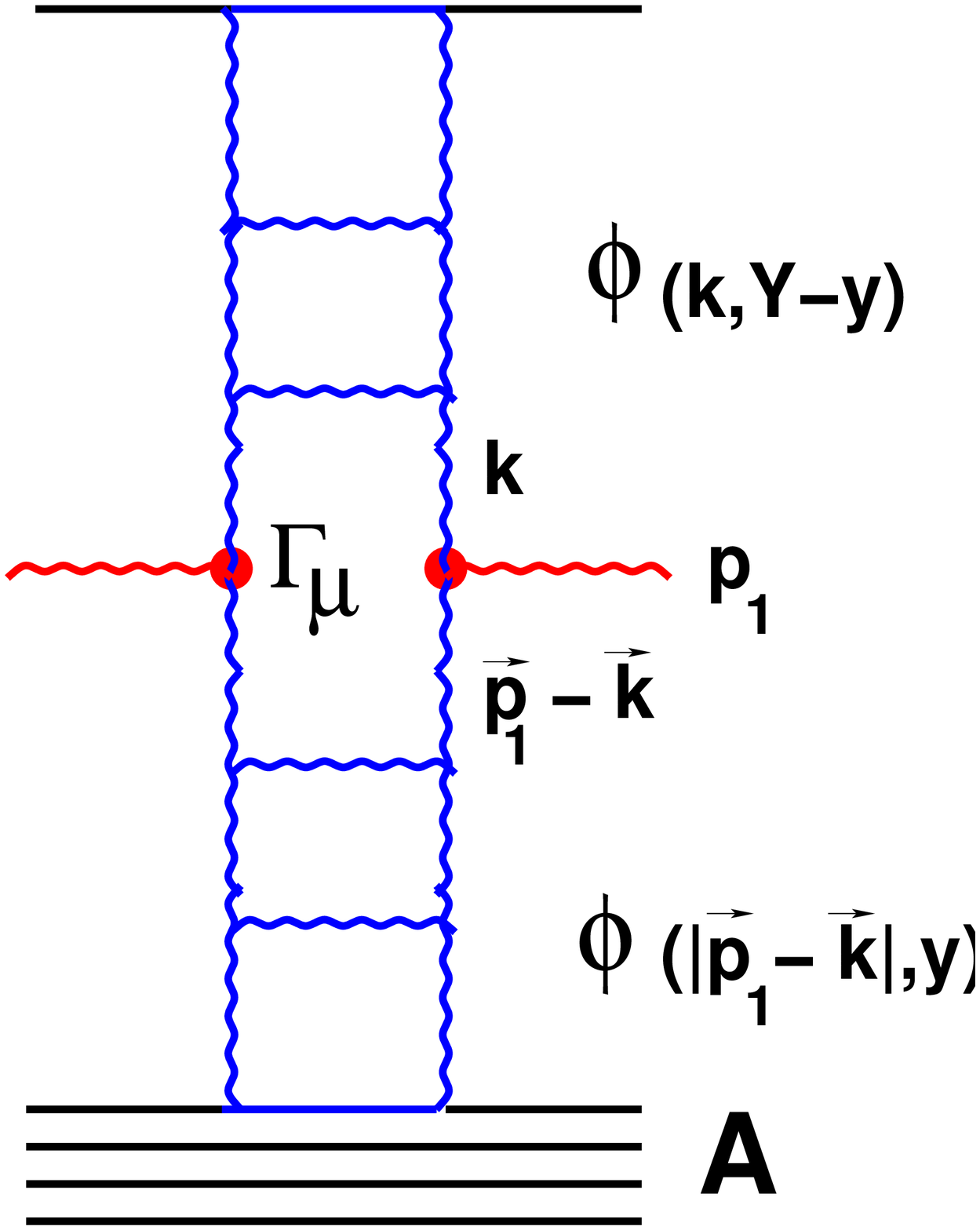,width=7cm,height=7.8cm}\\ 
\fig{cor2}-a & \fig{cor2}-b
\end{tabular} 
\caption{Double inclusive cross section for two correlated
jet production in the perturbative phase of QCD (\fig{cor2}-a) and single
inclusive cross section for a single jet production (\fig{cor2}-b).}
\label{cor2} 
\end{figure}

Due to AGK cutting rules \cite{AGK}, the contribution of the one parton
shower production to the double inclusive is described by one Mueller
diagram of \fig{cor2}-a.

The vertex of the gluon emission $\Gamma_\mu$ in \fig{cor2}-b (so-called
Lipatov vertex)  is equal to (see for example Ref. \cite{GLR}) 
\beq
\label{VG} \Gamma_\mu \,\,=\,\, \,\frac{2\,g}{p^2_{1t}}\,\left(p^2_{1,t}\,
k_{\mu}\,\,-\,k^2_t\,p_{1,\mu} \,\right) 
\eeq 
which leads to 
\beq
\label{VG2} \Gamma_\mu \,\Gamma^\mu\,\,=\,\,4\,\as
\,\frac{k^2_t\,(\vec{p}_{1,t} - \vec{k}_t)^2}{p^2_{1,t}} 
\eeq 
Taking these
equations into account one can see that the double inclusive cross section
given by the diagram of \fig{cor2}-a is equal to (see \fig{cor2}-a) (see
Ref.\cite{IF} for details)
 
$$\frac{1}{\sigma}\,\frac{d^2 \sigma}{d y_1\, d y_2\,d^2 p_{1,t}\,d^2 
p_{2,t}}\,\,= $$
$$
= \left(\frac{ 4 N_c\,\as}{N^2 
-1}\right)^2\,\,\frac{1}{p^2_{1,t}\,\,p^2_{2,t}}\,\,\int\,\,d^2\,k_t\,
\varphi_{projectile}(k^2_t,Y 
-y_1)\,\,\varphi_{target}( |\vec{p}_{1,t} - \vec{p}_{2,t} - 
\vec{k}_t|,y_2)$$
\beq  
\label{DIXS}  
=\,\,
\left(\frac{ 4 
N_c\,\as}{N^2-1}\right)^2\,\,\frac{1}{p^2_{1,t}\,\,p^2_{2,t}}\,\,F^{INCL}
(|\vec{p}_{1,t} - 
\vec{p}_{2,t}|,Y-y_1,y_2)\,\,
\eeq

It should be stressed that the single inclusive cross section of
\fig{cor2}-b can be rewritten through the same function $F^{INCL}$, namely
$$ \frac{1}{\sigma}\,\frac{d\sigma}{d y\,\,d^2 p_{1,t}}\,\,=\,\,
\left(\frac{ 4
N_c\,\as}{N^2-1}\right)\int\,\,d^2\,k_t\,\varphi_{projectile}(k^2_t,Y
-y)\,\,\varphi_{target}( |\vec{p}_{1,t} - \vec{k}_t|,y)$$ 
\beq
\label{SIXS} =\,\, \left(\frac{ 4
N_c\,\as}{N^2-1}\right)\,\,\frac{1}{p^2_{1,t}}\,\,F^{INCL}(p_{1,t},Y-y,y)\,\,
\eeq

\begin{figure}[h] 
\begin{center} 
\epsfysize=6cm 
\leavevmode 
\hbox{\epsffile{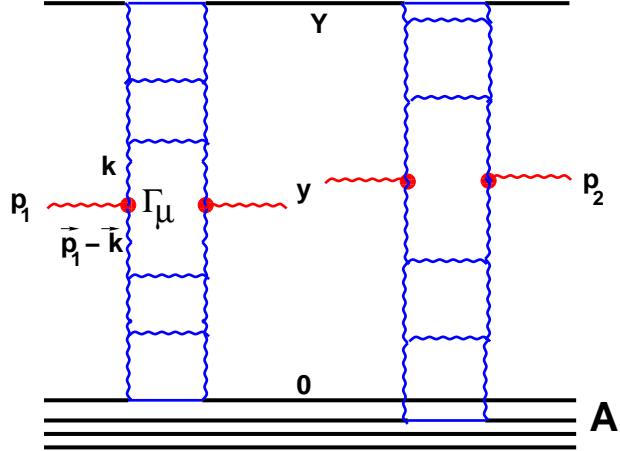}} 
\end{center} 
\caption{The Mueller diagram for the
production of two gluon jets from different parton showers. }
 \label{cor3}
\end{figure}

The production of jets from two different parton showers which is
described by the second diagram in \fig{jetcor} can be calculated using
the Mueller diagram of \fig{cor3}. It is easy to understand that this
diagram gives the double inclusive cross section in the factorized form
\beq 
\label{DXSIN} \frac{1}{\sigma}\,\frac{d^2 \sigma}{d y_1\, d y_2\,d^2
p_{1,t}\,d^2 p_{2,t}}\,\,= \,\,\,\frac{1}{\sigma}\,\frac{d\sigma}{d
y_1\,\,d^2 p_{1,t}}\,\times\,\frac{1}{\sigma}\,\frac{d\sigma}{d y_2\,\,d^2
p_{2,t}} 
\eeq 
corresponding to the independent (uncorrelated ) production
of two gluons with kinematic variables $(y_1, p_{1,t})$ and 
$(y_2,p_{2,t})$.

Let us now define the correlation function in the azimuthal angle $\phi$
between the two gluons as a probability to find a second gluon with
rapidity $y_2$ and transverse momentum $p_{2,t}$ moving at the angle
$\phi$ with respect to the trigger gluon with rapidity $y_1$ and
transverse momentum $p_{1,t}$. Defined this way correlation function has
the following form: 
\beq 
\label{CRF} R(\phi; p_1,p_2,y_1=y_2)\,\,= 
\eeq
$$\frac{F^{INCL}(|\vec{p}_{1,t} -
\vec{p}_{2,t}|,Y-y_1,y_2)\,\,+\,\,F^{INCL}(p_{1,t},Y-y_1,y_1)\,\,
F^{INCL}(p_{2,t},Y-y_2,y_2)}{\int\,\,d\,\phi\,\left(F^{INCL}(|\vec{p}_{1,t}
- \vec{p}_{2,t}|,Y-y_1,y_2)\,\,+\,\,F^{INCL}(p_{1,t},Y-y_1,y_1)\,\,
F^{INCL}(p_{2,t},Y-y_2,y_2)\right)}. $$

The azimuthal angle dependence originates only from the production of two
gluon jets from the same parton shower (see \eq{DIXS}) while the second
mechanism (see the secoond diagram in \fig{jetcor}) leads to the constant
background (see \eq{DXSIN}).

The shape of the differential cross sections \eq{DIXS} and \eq{DXSIN}
depends crucially on the unintegrated gluon densities $\varphi$.  Here we
will use a simple model for these functions adopted earlier in Ref.
\cite{KL} : \beq \label{PHISAT}
\varphi(x;p^2_t)\,\,=\,\,\left\{\begin{array}{l}\,\,\,\,
\frac{\kappa}{\as(Q^2_s)}\, S\,\,(1\, -\, x)^4\,\,\,\,p_t\,<\,Q_s(x)\,\,;
\\ \\ \,\,\,\, \frac{\kappa}{\as(Q^2_s)}\,
S\,\frac{Q^2_s(x)}{p^2_t}\,\,(1\, -\, x)^4\,\,\,\,p_t\,>\,Q_s(x)\,\,;
\end{array} \right. \eeq

In \eq{PHISAT} we neglect, for the time being, the anomalous dimension of
the gluon densities and use a very simplified assumption about the
behavior of $\varphi$ reflecting the fact that inside the saturation
region the density is large and changes slowly \cite{MV1}.  The numerical
factor $\kappa$ can be found from RHIC data on ion-ion collisions, but the
value of $R$ (see \eq{CRF}) does not depend on it. We introduce, as before
\cite{KN,KL,KLN,KLM,KLNDA}, the factor $(1 \,-\,x)^4$ which describes that
the gluon density is power suppressed at $x\,\rightarrow\,1$ according to
the quark counting rules. However if we restrict ourselves to calculation
of the correlation function at $y_1=y_2 = 0$ then the influence of these
factors is very small. In general, one can expect this ansatz for
$\varphi$ of \eq{PHISAT} to be a rather crude model but it turns out to be
quite successful in describing the data on rapidity and transverse
momentum distributions \cite{SZC}. Therefore, we hope that our
calculations will provide a reasonable guideline for the experimental
measurements.

\begin{figure}[h] \begin{tabular}{c c} 
\epsfig{file=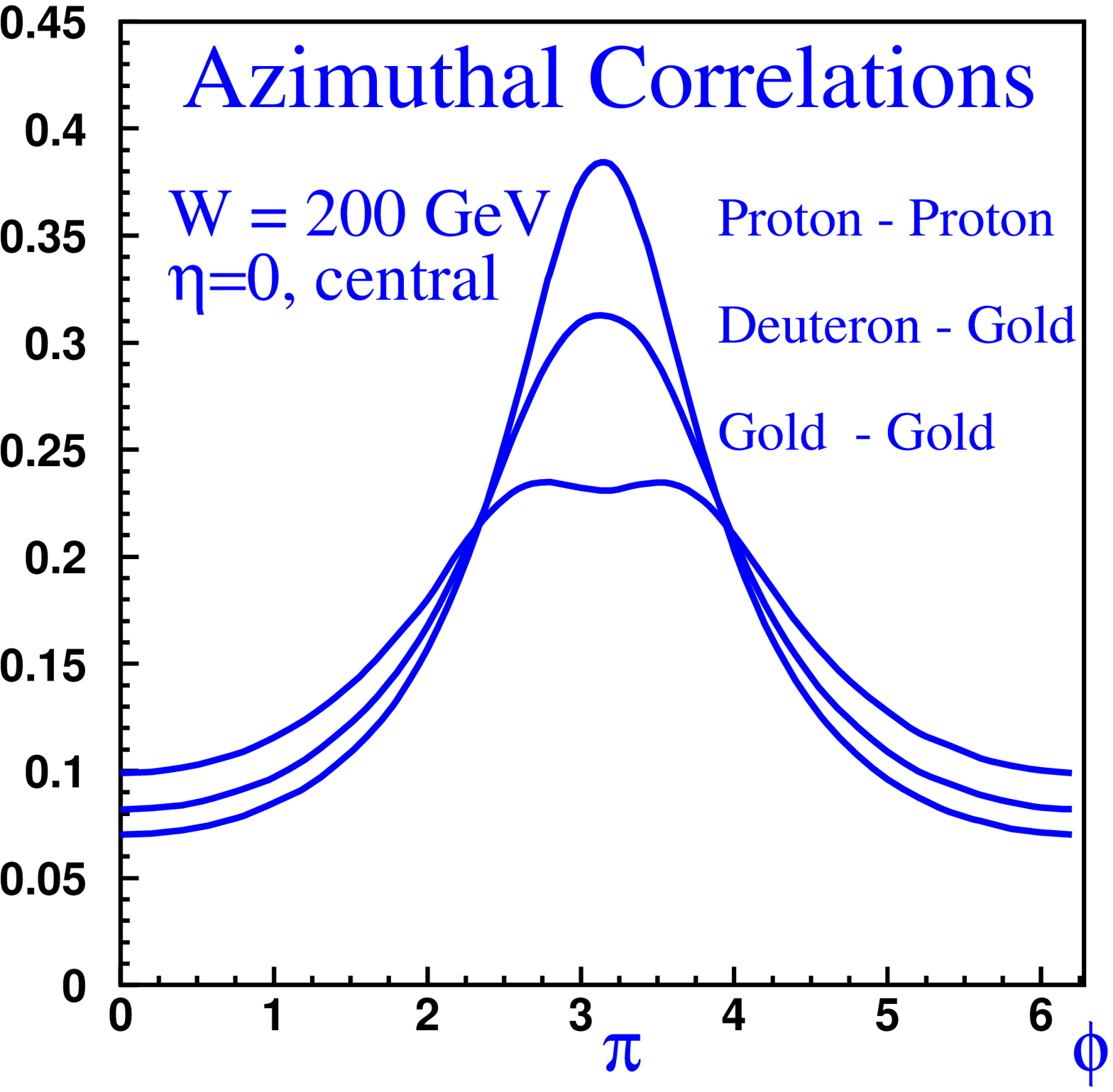,width=80mm} &
\epsfig{file=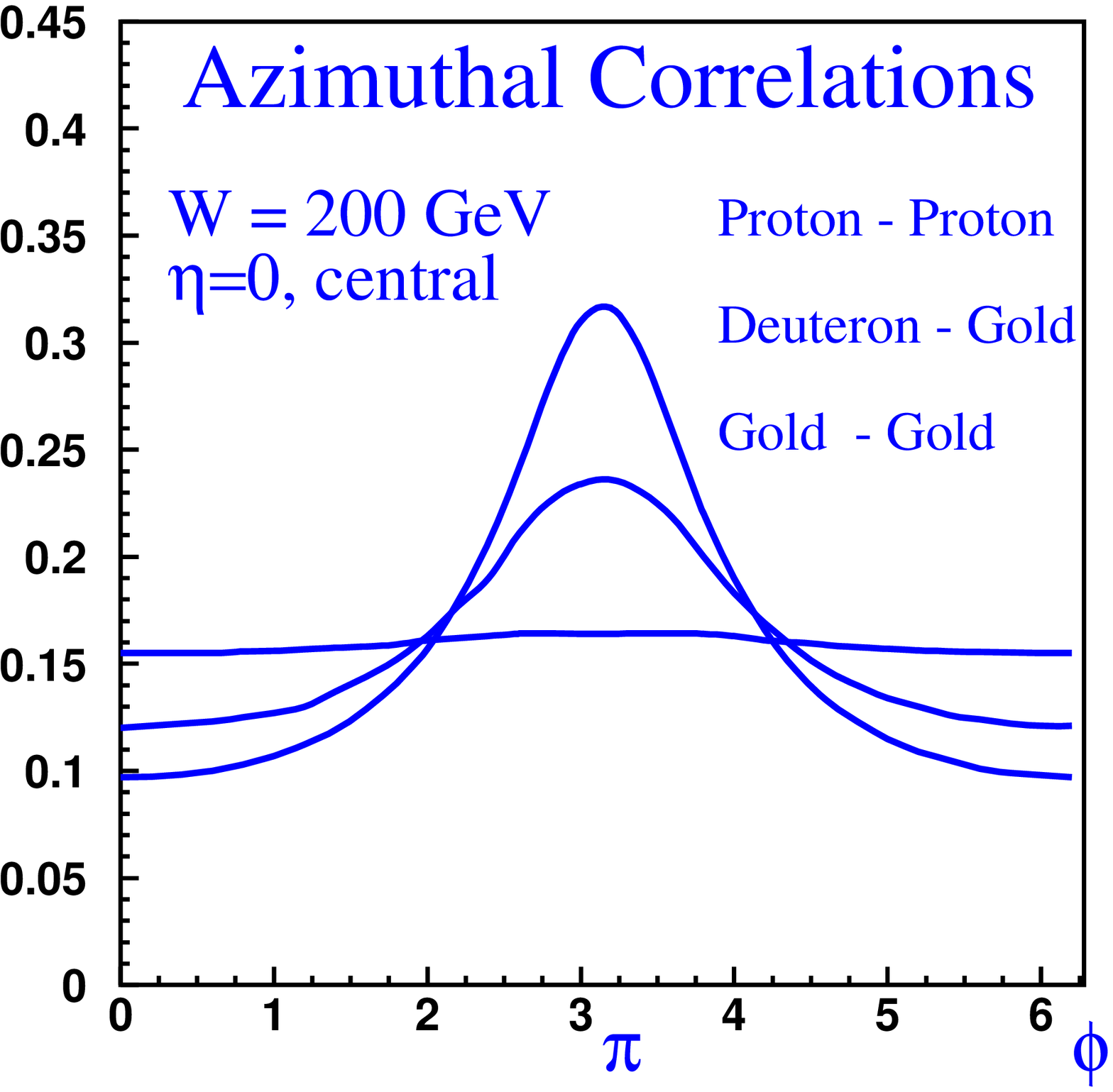,width=80mm} \\ \fig{cor1ps}-a &
\fig{cor1ps}-b \\ \end{tabular} \caption{ \fig{cor1ps}-a :  Two gluon jet
production in one parton shower normalized in a such way that the integral
over azimuthal angle is equal to 1. The upper curve corresponds to
proton-proton interaction, the middle one describes the deuteron-gold
interaction while the third curve is related to gold-gold interaction. The
calculation is performed for two jets with $p_1 = 4\,GeV$ and $p_2 =
2\,GeV$ which corresponds to STAR experiment measurements \cite{STAR}.
\fig{cor1ps}-b: Same as in \fig{cor1ps}-a, but for the sum of one and two
parton showers.
  }
 \label{cor1ps}
\end{figure}

For the numerical estimate we take the value of saturation scale
$Q^2_s(A;y=0) = 2.44\,GeV^2$ for gold at W=200 GeV and $Q^2_s(p;y=0) =
0.26\,GeV^2$ for the proton in accord with our multiplicity calculations for
deuteron - gold collisions \cite{KLNDA}. The result is plotted in
\fig{cor1ps}.

One can see that the azimuthal angle distribution has a maximum at $\phi =
\pi$ which corresponds to the jet produced in the opposite direction to
the trigger jet. The width of the angular distribution is different for
$pp$ and $dA$ processes. Introducing the Gaussian distribution \beq
\label{GD} R =\frac{1}{\sqrt{2 \,\pi} \sigma}\,\,e^{ - \frac{ ( \phi -
\pi)^2}{ 2\,\sigma^2}} \eeq with the width $\sigma$, we can see from
\fig{cor1ps} that for proton-proton collision $\sigma\,=\,0.7$ while for
deuteron-gold interaction $\sigma\,=\,0.8$. Therefore, the difference
between the $pp$ and $dA$ processes at mid-rapidity appears quite small.
However, for gold-gold interactions the $\phi$ distribution in
\fig{cor1ps} appears quite different from the Gaussian given by \eq{GD}.
If fitted to the Gaussian form, the value of $\sigma$ appears
approximately $\sigma = 1.0 \div 1.2$; however this does not characterize
well the distribution. This result is easy to understand qualitatively,
since for large $p_1$ and $p_2$ (so large that $| p_1 - p_2| \,> Q_S(A)$ )
the integral over $k$ in \eq{DIXS} stems from the region where $k
\,\approx \,Q_s(p) < | p_1 - p_2|$ and 
\beq 
\label{WEST}
F^{INCL}(|\vec{p}_{1,t} -
\vec{p}_{2,t}|,Y-y_1,y_2)\,\,\propto\,\,\frac{Q^2_s(A)}{(p_1 - p_2)^2
\,+\,2p_1\,p_2 \,cos \phi}\,\,\approx\,\,\frac{ \frac{Q^2_s(A)}{(p_1 -
p_2)^2}}{1 + \frac{2p_1\,p_2}{(p_1 - p_2)^2}\,\frac{(\phi -\pi)^2}{2}}
\eeq

In this formula the typical width of the azimuthal angle distribution is
equal to $\sigma_{typical} \,=\,(p_1 - p_2)^2/2\,p_1\,p_2$. For the
kinematics used in the STAR experiment \cite{STAR}, $p_1 = 4\,GeV $ and
$p_2 = 2\,GeV$, we obtain $\sigma_{typical} \,\approx 0.5$ from this
simple formula. As was mentioned above, the fact that \fig{cor1ps} leads
to a large value of $\sigma_{typical}$ is due to the region of integration
over $ k \approx Q_s(A)$.

In \fig{cor1ps}-b we plot $R(\phi; p_1,p_2,y_1=y_2)$ for $p_1 = 4\, GeV $
and $p_2 = 2\, GeV$ as a function of $\phi$. $R$ has a meaning of the
probability to find a particle with momentum $p_2$ at a definite angle
$\phi$ if a trigger particle has a momentum $p_1$. The curves in this
figure are quite different from \fig{cor1ps} due to the background from
the production of gluon jets from two parton showers. This background is
proportional to $N_{part}^2$ where $N_{part}$ is the number of
participants (see Refs. \cite{KL,KLM} ). In fact, this background becomes
so large for gold-gold collision that we cannot see the azimuthal
correlations. However, for proton - proton and deutron-gold collisions the
background is not so large and we see a clear peak at $\phi =\pi$.

\begin{figure} 
\begin{center} 
\epsfysize=11cm 
\leavevmode 
\hbox{\epsffile{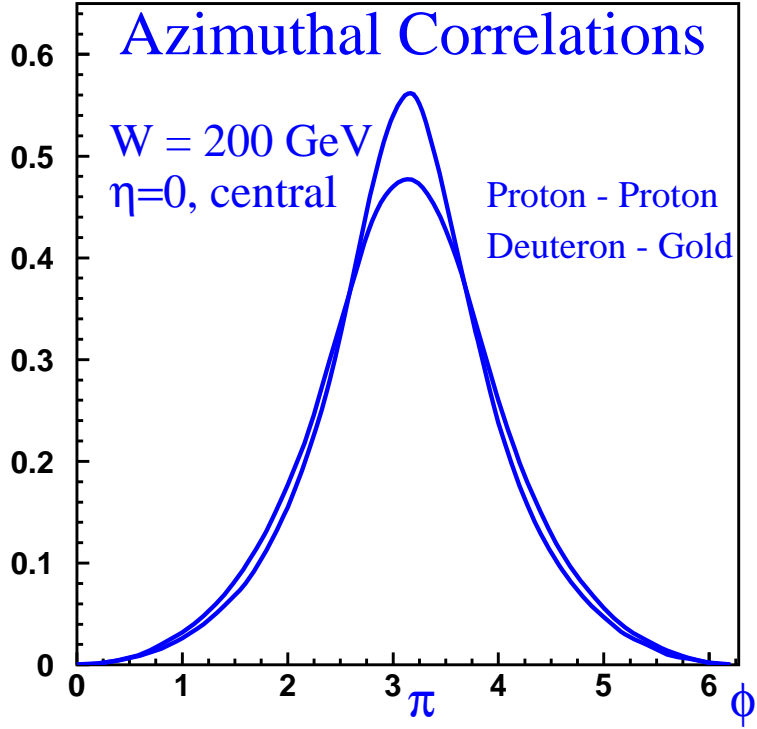}} 
\end{center} 
\caption{The azimuthal angle
correlations after substraction of the background. The upper curve
corresponds to proton-proton interaction while the lower one describes
the deuteron-gold interaction. The calculation is performed for two jets
with $p_1 = 4\,GeV$ and $p_2 = 2\,GeV$ which corresponds to STAR
experiment measurements \cite{STAR} }
 \label{corsub}
\end{figure}

However, experimentalists in their analysis sometimes substract the flat
background -- the "pedestal" \cite{STAR}. Doing such a subtraction for
\fig{cor1ps}-b and normalizing the $\phi$ - distribution to the unit area
we arrive at the $\phi$ correlation functions shown in \fig{corsub}. It
should be stressed that the ratio of the areas is equal to pp/pA= 1.16.
Thus one can see that we get quite similar $\phi$ - distributions for
proton-proton collisions and for deuteron-gold ones, in (at least
qualitative) agreement with the data \cite{STAR}.

To make the main features of our approach even more transparent, let us
use the following simple model for the unintegrated structure function
$\varphi$ which allows us to make all calculations analytically:  \beq
\label{PHISATSM} \varphi(x;p^2_t)\,\,=\,\,\frac{\kappa\,S}{\as(Q^2_s)}\,(1
- x)^4 \,\frac{Q^2_s}{p^2_t\,+\,Q^2_s} \eeq One can see that for $p_t
\,\ll\,Q_s$ and for $p_t \,\gg\,Q_s$ we have the function $\varphi$ of
\eq{PHISAT} , while in the region of $p_t \,\approx Q_s$ $\varphi$ of
\eq{PHISATSM} is different. The advantage of \eq{PHISATSM} is the fact
that we can calculate $F^{INCL}$ of \eq{DIXS} and \eq{SIXS} analytically.
Indeed, introducing Feynman parameters and taking integral over $k_t$ in
\eq{DIXS} and \eq{SIXS} we obtain 
\begin{eqnarray}
F^{INCL}(p_t,Q_s,q_s)\,\,&=&\,\,\pi\,\int^1_0\,dt\,\frac{1}{t\,(1
-t)\,p^2_t\,\,+\,\,t\,(Q^2_s - q^2_s)\,\,+\,\,q^2_s}\,\,\label{SMIN}\\
 &=&\,\,\frac{\pi}{2\,h(p_t,Q_s,q_s}\,\,\ln 
\left(\frac{p^2_t 
\,+\,Q^2_s\,+\,q^2_s\, 
+\,h(p_t,Q_s,q_s)}{p^2_t \,+\,Q^2_s\,+\,q^2_s\,
-\,h(p_t,Q_s,q_s)} \right) \nonumber
\end{eqnarray}
where $Q_s$ and $q_s$ are saturation momenta of projectile and target and 
\beq 
\label{h}
h(p_t,Q_s,q_s)\,\,=\,\,\sqrt{(p^2_t + (Q_s - q_s)^2\,(p^2_t + (Q_s + q_s)^2}
\eeq

\begin{figure}[h] 
\begin{tabular}{c c} 
\epsfig{file=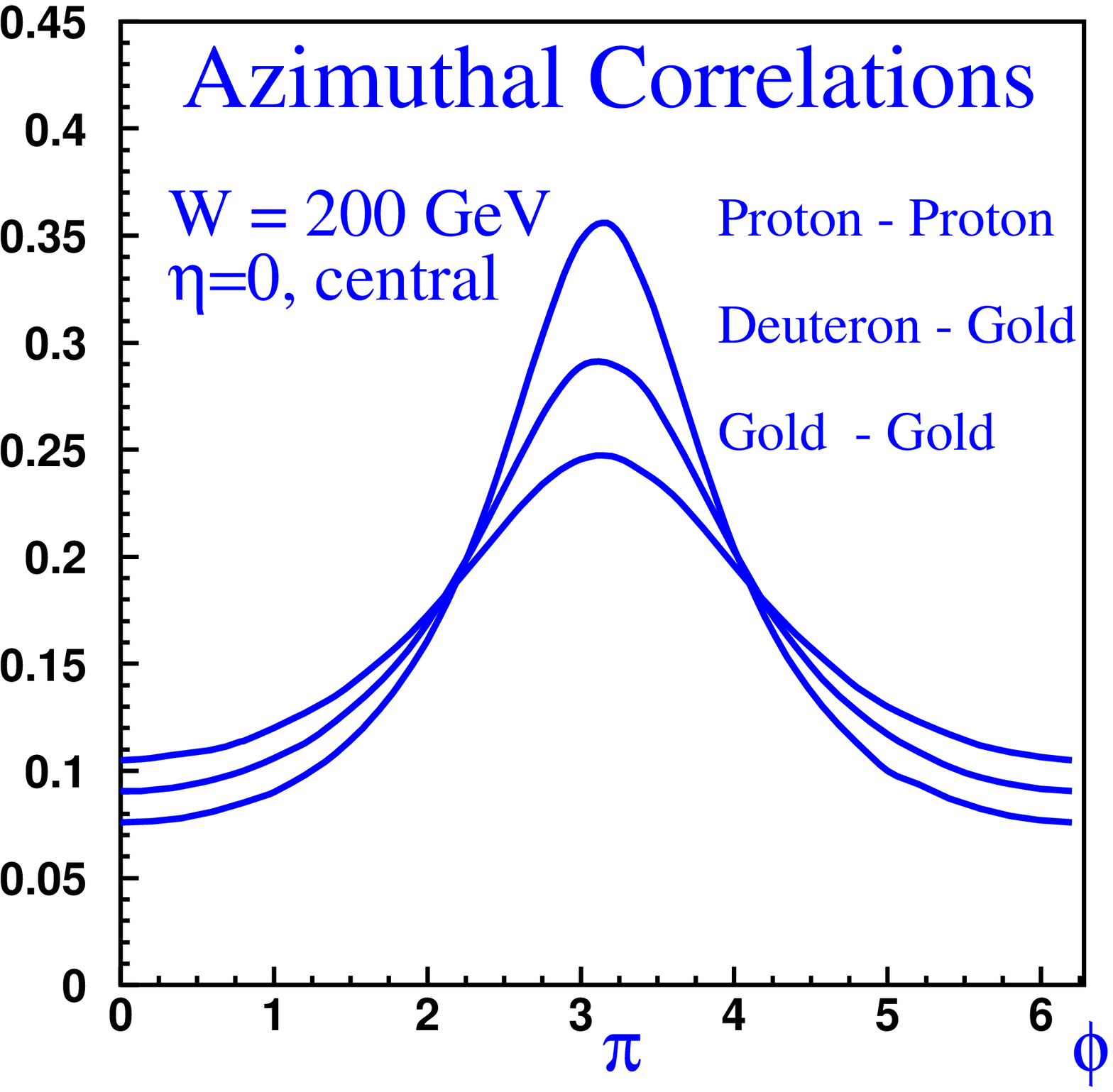,width=80mm} &
\epsfig{file=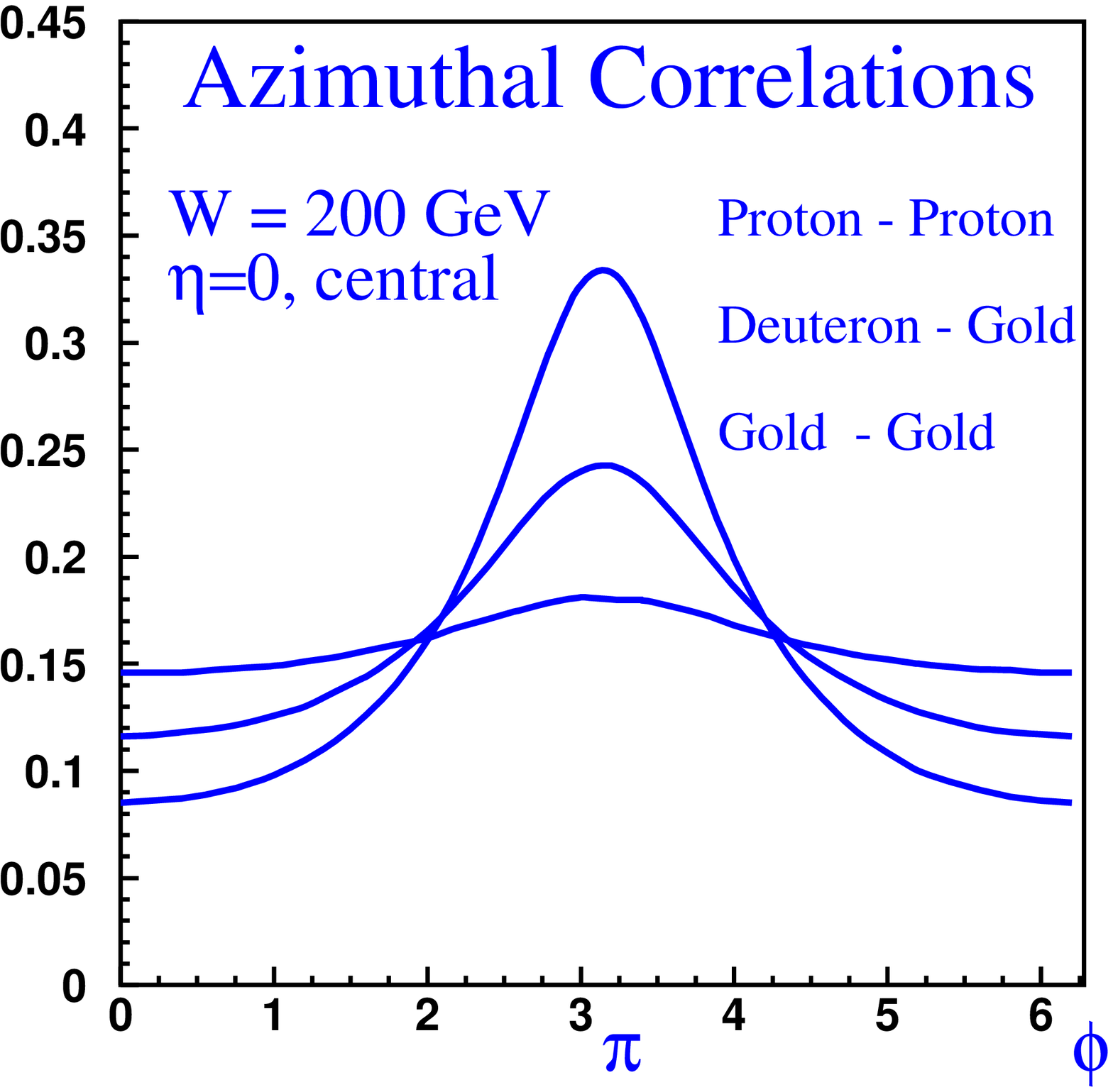,width=80mm} \\ 
\fig{corsm1ps}-a &
\fig{corsm1ps}-b \\ 
\end{tabular} 
\caption{\fig{corsm1ps}-a: Two gluon jet
production calculated in the simple model (see \eq{SMIN}) for the
unintegrated stricture function $\varphi$ given by \eq{PHISATSM} in one
parton shower normalized in a such way that the integral over azimuthal
angle is equal to 1. The upper curve corresponds to proton-proton
interaction, the middle one describes the deuteron-gold interaction while
the third curve is related to gold-gold interaction. The calculation is
performed for two jets with $p_1 = 4\,GeV$ and $p_2 = 2\,GeV$ which
corresponds to STAR experiment measurements \cite{STAR}. \fig{corsm1ps}-b:
Two gluon jet production, calculated using \eq{SMIN}, in one and in two
parton showers normalized in a such way that the integral over azimuthal
angle is equal to 1.} 
\label{corsm1ps} 
\end{figure}

\begin{figure}[h] 
\begin{tabular}{c c} 
\epsfig{file=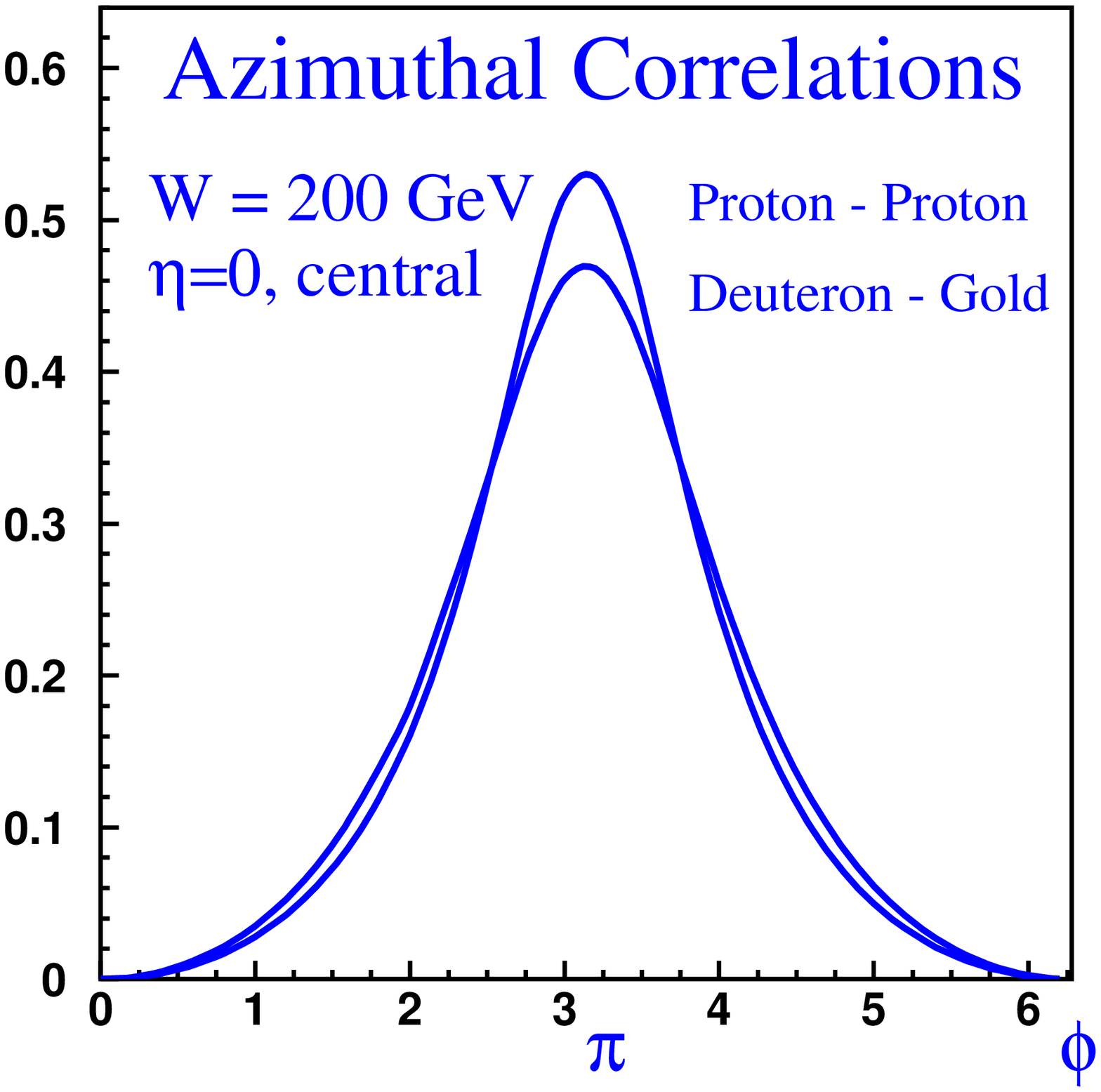,width=80mm} &
\epsfig{file=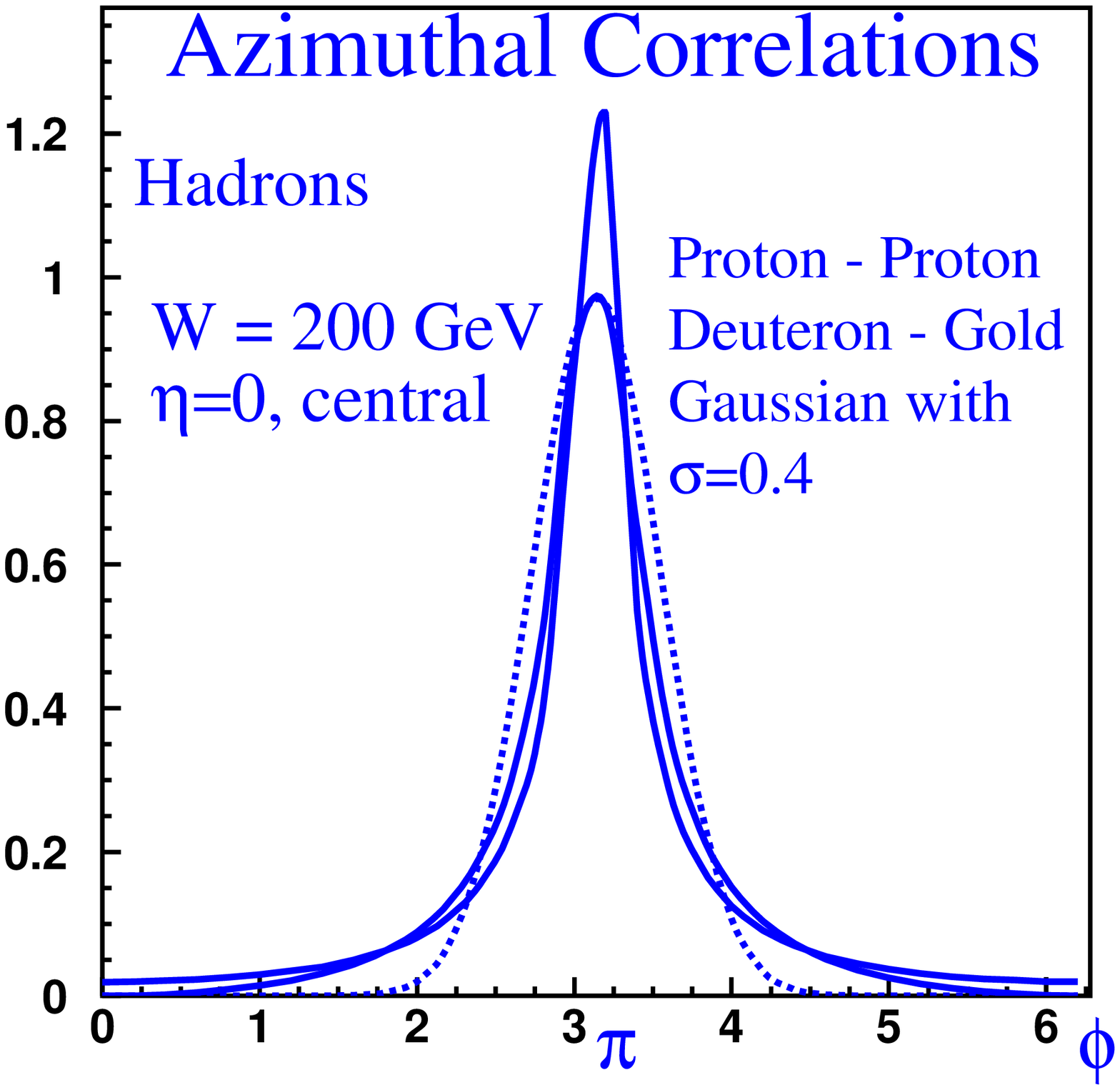,width=80mm} \\ 
\fig{corsmsub}-a &
\fig{corsmsub}-b \\ 
\end{tabular} 
\caption{\fig{corsmsub}-a:  The
azimuthal angle correlations, calculated using \eq{SMIN}, after
subtraction of the background. The upper curve corresponds to
proton-proton interaction while the lower one  describes the
deuteron-gold interaction. The calculation is performed for two jets with
$p_1 = 4\,GeV$ and $p_2 = 2\,GeV$ which corresponds to STAR experiment
measurements \cite{STAR} \fig{corsmsub}-b: The azimuthal angle
correlations of produced hadrons after substraction of the background.  
The upper curve corresponds to proton-proton interaction while the lower
one  describes the deuteron-gold interaction. The calculation is
performed for two jets with $p_1 = 4\,GeV$ and $p_2 = 2\,GeV$ which
corresponds to STAR experiment measurements \cite{STAR}. The dotted line
is the Gaussian distribution of \eq{GD} with $\sigma = 04$. Fragmentation
is included as described in the text. }
\label{corsmsub} \end{figure}

One can see that the two scales governing the azimuthal angle dependence
emerge:  
\beq 
\label{SMW} 
\sigma^{(-)}_{typical}\,\,=\,\frac{(p_1 - p_2)^2
+ (Q_s -q_s)^2}{2\,p_1\,p_2}\,\,;\,\,\,\,\,\,\,\,\,\,\,\,
\sigma^{(+)}_{typical}\,\,=\,\frac{(p_1- p_2)^2 + (Q_s + q_s)^2}{2\,p_1\,p_2}
\,\,; 
\eeq 
In the case of ion-ion
collisions the value of $\sigma^{(+)}_{typical}$ is rather large, and this
leads to a wide distribution in $\phi$. Using this simple model, we
recalculate the results presented in \fig{cor1ps}, \fig{cor1ps}-b and
\fig{corsub} using \eq{SMIN}; the results are presented in \fig{corsm1ps},
\fig{corsm1ps}-b and \fig{corsmsub}.

Examination of these results leads us to the conclusion that our approach
explains, at least on a qualitative level, two of the most striking
experimental facts: (i)  the strong suppression of the azimuthal
back-to-back correlations in gold-gold collisions and (ii) the close
similarity of these correlations for the proton-proton and deuteron-gold
interactions. (The final-state interactions of the jets in ion-ion
collisions are expected to broaden the observed azimuthal correlations
even more.)

It is interesting to see if we can reproduce the measured widths of the
$\phi$-diastribution quantitatively.  As one can see from \fig{corsub},
the calculated width is larger than the experimental one. One obvious
reason which can contribute to this difference is the fragmentation of the
jets: indeed, we have computed the correlation function for gluons, not
for the measured hadrons.  As a result of fragmentation, the fall-off of
the inclusive cross section (and also of $F^{INCL}$ in \eq{DIXS} and
\eq{DXSIN})  with the transverse momentum becomes more steep. To take
account of this, we use the fragmentation function of gluon jet to hadrons
from Ref. \cite{KKP} . One can see that the resulting width of the
distribution in \fig{corsmsub}-b is indeed much smaller than in
\fig{corsub}, and is now close to the experimental one. One can also see
that the real distribution is not quite Gaussian.

\begin{figure} 
\begin{center} 
\epsfysize=11cm 
\leavevmode 
\hbox{\epsffile{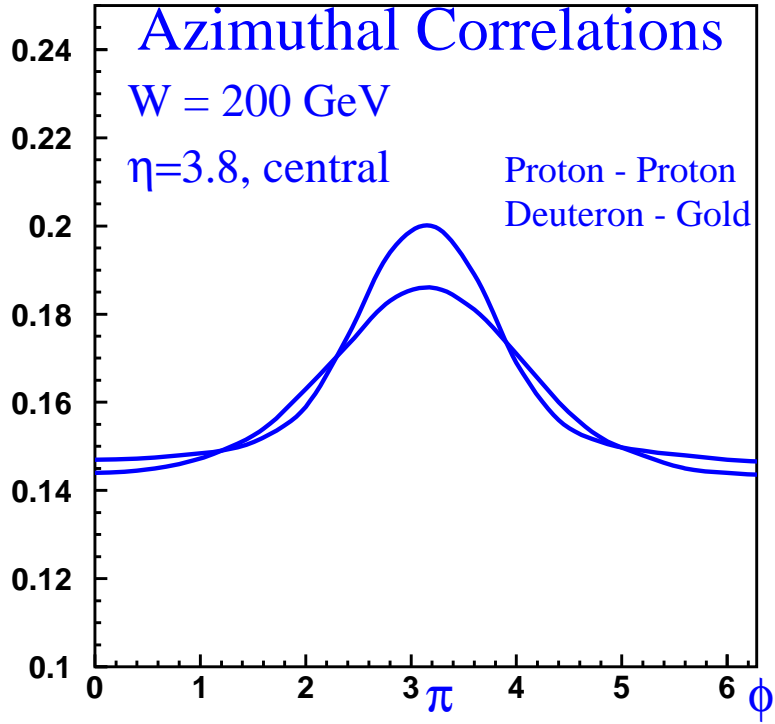}} 
\end{center} 
\caption{The azimuthal angle
correlations of produced jets at forward direction with rapidity $\eta
=3.8$. The value of the trigger momentum is taken $1.5\,GeV$ while the
hadrons produced in the backward direction were integrated over transverse
momenta from $0.2\,GeV$ to $1.5 \,GeV$.  }
\label{corfor}
\end{figure}

One of the possibilities to check the validity of our approach is to
measure the azimuthal correlations in the kinematics when both of the
detected hadrons are measured at forward rapidity. Since the nuclear
saturation momentum increases with rapidity, $Q_s(A,y)=
Q_s(A,y=0)\,\exp[\lambda\,y]$ with $\lambda \,\approx\,0.25 \,\div \,0.3$
\cite{GW}, we expect a wider distribution in the azimuthal angle,
than at $y = 0$.  \fig{corfor} shows our prediction for deuteron-gold and
proton - proton scattering for $y =3.8$.

\begin{figure}
\begin{minipage}{10cm}
\begin{center}
\epsfxsize=9cm
\epsfxsize=8cm
\leavevmode
\hbox{ \epsffile{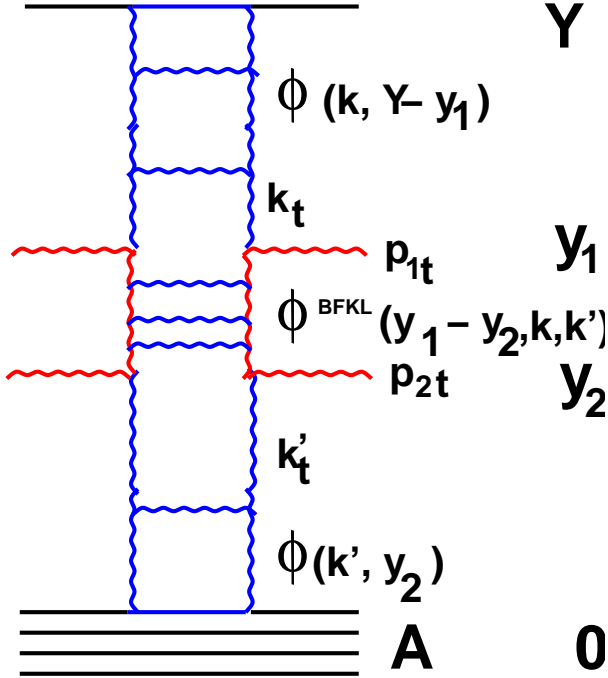}}
\end{center}
\end{minipage}
\begin{minipage}{6cm}
\caption{The BFKL emission in the double inclusive cross section. }
 \label{cor0yfor}
\end{minipage}
\end{figure}

A most interesting opportunity to investigate the CGC dynamics in the
quantum domain is to study reactions \cite{Les} in which the trigger
hadron with transverse momentum $p_1$ is generated in the forward
direction ( say, $y_1 = 3.8$) while the recoiling particle(s) produced at
the azimuthal angle $\phi$ is detected in the central rapidity region
($y_2 = 0$, see \fig{cor0yfor}). The interest in this kinematics stems
from the fact that a large rapidity interval $\Delta y = y_1 - y_2$
between the detected particles enhances the effects of quantum evolution,
since the probability of gluon emission is proportional to $ \alpha_s \
\Delta y$. In fact, the study of gluon jets separated in rapidity has been
proposed by Mueller and Navelet \cite{MuellerN} as a way to investigate
the properties of BFKL evolution.  In the case of nuclear target, the
quantum evolution enhances the influence of the saturation and extends it
to larger transverse momenta.

The double inclusive cross section for such  kinematics can be written in
the form: 
\beq 
\label{DCF} 
\frac{1}{\sigma}\,\frac{d\sigma}{d y_1\,d\,y_2
\,\,d^2 p_{1,t}\,d^2p_{t,2}}\,\,=\,\,\left(\frac{ 4
N_c\,\as}{N^2-1}\right)^2\,\frac{1}{p^2_{1,t}}\,\frac{1}{p^2_{2,t}} \times 
\eeq 
$$ 
\int\,\,d^2\,k_t\,,d^2\,k'_t
\varphi_{projectile}\left(k^2_t,Y
-y_1\right)\,\, \varphi^{BFKL}\left(y_1 -
y_2,k_t,k'_t;\phi\right)\,\,\varphi_{target}\left(k'_t,y_2\right) 
$$ 
where
$\varphi^{BFKL}$ is the BFKL scattering amplitude for two gluons. This
amplitude takes into account the emission of gluons with rapidities
between $y_1$ and $y_2$ (see \fig{cor0yfor}). This amplitude can be
written as series \cite{BFKL} 
\beq 
\label{SERIES}
\varphi^{BFKL}\left(y_1-y_2,k_t,k'_t;\phi\right)\,\,=\sum^{\infty}_{n = 0}
\,\,Cos( n \,\phi)\,\,\varphi^{BFKL}_n \left(y_2 - y_1,k_t,k'_t\right)
\eeq where $\phi$ is the azimuthal angle and $\varphi^{BFKL}_n$ are the
eigenfunctions of the BFKL equation, which at high energies behave as \beq
\label{PHIN} \varphi^{BFKL}_n \left(y_2
-y_1,k_t,k'_t\right)\,\,\,\rightarrow\,\,\,e^{\omega_{n}\,\,( y_2 - y_1)}
\eeq 
where 
\beq 
\label{OMEGAN} 
\omega_n \,\,\,=\,\,\,\frac{2\,\alpha_S
N_c}{\pi}\,\left(\,\psi(1)\,\,-\,\,\psi( \frac{1}{2} + n ) \,\right) 
\eeq
However, in \eq{SERIES} only the first term has a positive intercept
($\omega_0 >0 $) while all other terms fall off as a function of energy.
Indeed, \eq{OMEGAN} gives 
$$ 
\omega_0 \,\,=\,\,\frac{4\,\ln2\,\alpha_S
N_c}{\pi}\,\,\,\,\,\,; \,\,\,\,\,\omega_1 \,\,=\,\,-\,\frac{1.22\,\alpha_S
N_c}{\pi}\,\,\,\,\,\,; \,\,\,\,\,\omega_2
\,\,=\,\,-\,\,\frac{2.56\,\alpha_S N_c}{\pi} 
$$ 
We thus replace
$\varphi^{BFKL}$ by sum of two first terms with $n=0$ and $n=1$, since
other terms are supressed at large value of $|y_1 - y_2|.$

\begin{figure}[h] 
\begin{tabular}{c c} 
\epsfig{file=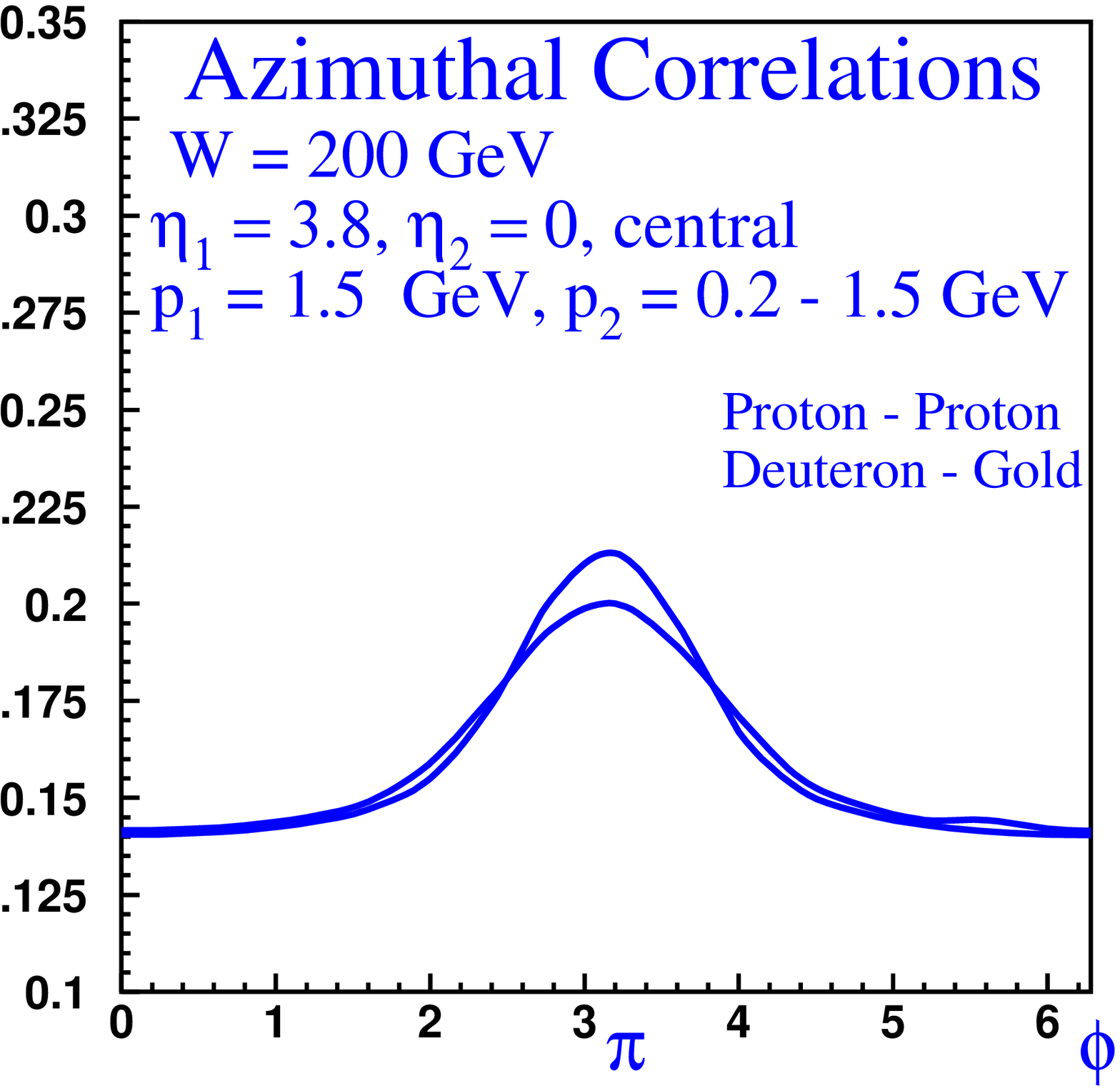,width=80mm} &
\epsfig{file=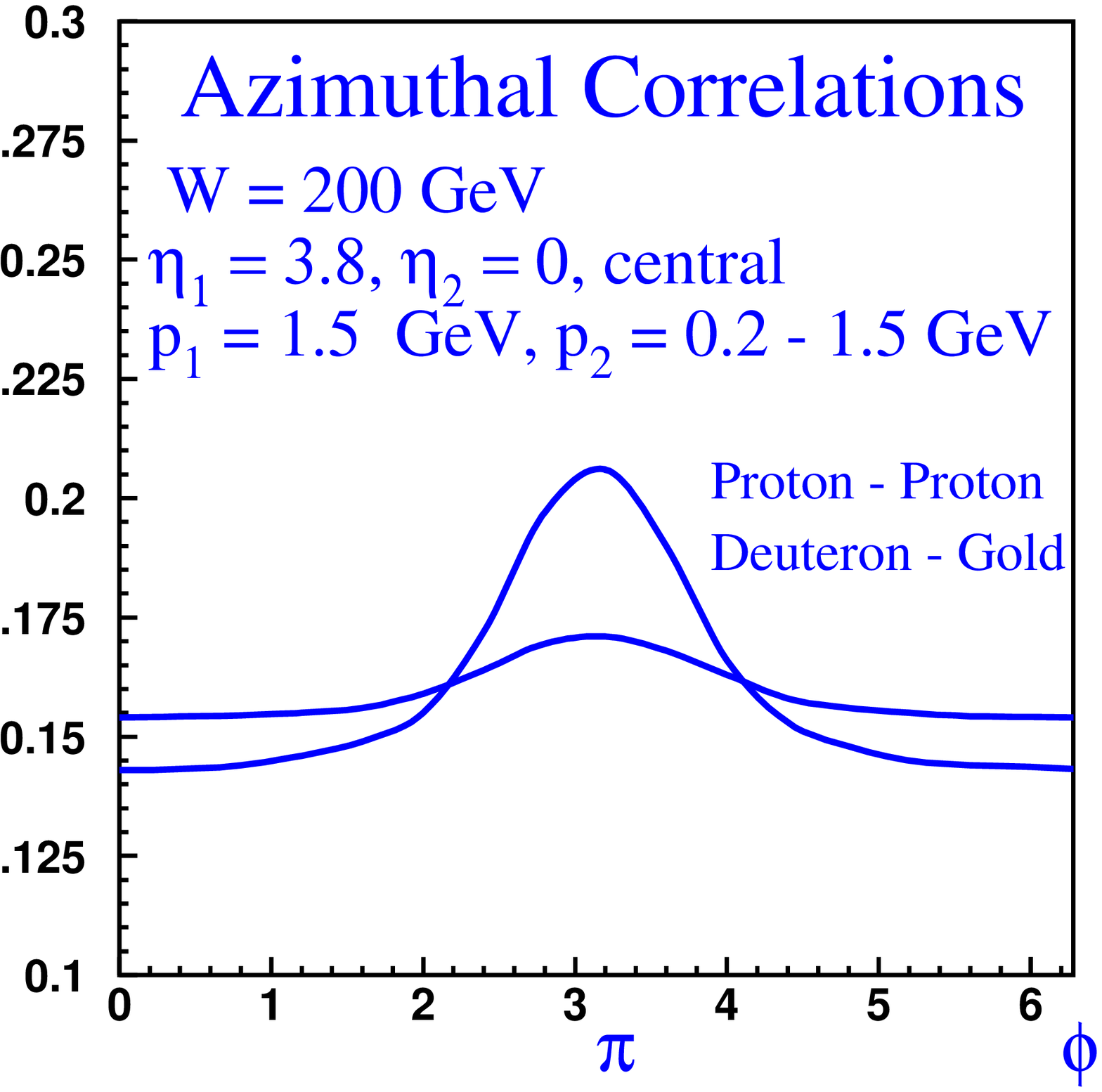,width=80mm} \\ 
\fig{corfor-i}-a &
\fig{corfor-i}-b \\ 
\end{tabular} 
\caption{\fig{corfor-i}-a:  The
azimuthal angle correlations of produced jets {\it in one parton shower}
at forward direction with rapidity $\eta_2 =0$ and with
$1.5\,GeV\,>\,p_2\,>\,0.4\,GeV$ when the trigger jet is at $\eta_1 = 3.8$
and with transverse momentum $p_1 = 1.5\,GeV$.  \fig{corfor-i}-b: The
azimuthal angle correlations of produced jets at forward direction with
rapidity $\eta_2 =0$ and with $1.5\,GeV\,>\,p_2\,>\,0.4\,GeV$ when the
trigger jet is at $\eta_1 = 3.8$ and with transverse momentum $p_1 =
1.5\,GeV$. Both, production from one parton shower and from two parton
showers are taken into account. } 
\label{corfor-i} 
\end{figure}

\fig{corfor-i} shows the normalized azimuthal distribution of the jets
when the recoiling jet has $\eta_2 = 0$ and $0.2 \,GeV \,\,<\,p_2\,<
1.5\,GeV$ while the trigger jet has $\eta_1 = 3.8$ and $p_1 = 1.5\,GeV$
{\it assuming that both of the jets are produced from one parton shower}
(see \fig{cor2}-a and \fig{cor0yfor}). One can see that in both cases (for
proton - proton and deuteron-gold collisions ) we expect sufficiently
strong correlations. For deuteron-gold collision the width of the
distribution in the azimuthal angle $\phi$ is only  30\% larger than for
the proton -proton scattering.

This result is not final though, since the main difference between the two
cases is in the independent production from two parton
showers ( see \fig{cor3} ).  In the CGC phase we have a saturation of the
parton densities which is expressed in our assumption for the functions
$\varphi$'s (see \eq{PHISAT}). In this region the diagrams for independent
two jet production (see \fig{cor3} ) are much more important than the
production of two jets from the single parton shower. Indeed, diagrams of
\fig{cor2}-a and \fig{cor0yfor} lead to the cross section of the order of
1 while the independent production given by \fig{cor3} leads to the cross
section of the order of $ ( 1/\alpha_S(Q_s))^2$. In reality,
\fig{corfor-i}-b shows that the azimuthal correlation in deuteron - gold
collision leads to a maximum around $\phi = \pi$ which the height of only
17\%.  For proton -proton we still expect a sizable effect of around 57\%.
One can see therefore that the quantum evolution effects in the CGC indeed
induce a large difference between the back-to-back correlations expected
in $pp$ and $dA$ collisions.

Let us now discuss the main uncertainties involved in our calculations.
Our estimates of the independent production in comparison with the
production in one parton shower have large errors because they involve the
normalization of the inclusive cross section.  The relative contribution
of the diagram of \fig{cor3} to the diagram of \fig{cor2}-a (we define $Z$
as a numerical factor which should stand in front of the diagram of
\fig{cor3} which is defined as the square of expression given by
\eq{SIXS}.)  can be expressed through the normalization constant that we
have introduced earlier in describing the multiplicity in d A collisions
\cite{KLNDA}. Denoting this constant by $C$ we have for the normalization
of the relative contribution ($Z$) the following expression 
\beq
\label{NORM} Z\,\,=\,\,\frac{(N^2_c - 1) \cdot C \cdot 4 \pi}{9
\,\alpha_S(Q_s) \,n_{jet}} 
\eeq 
where $n_{jet}$ is the hadron multiplicity
of jet with transverse momentum $Q_s$.  The uncertainty in calculation of
$n_{jet}$ is large and in our numerical estimates we take $n_{jet}= 1.5$.  
For proton -proton collisions we introduce an additional normalization
factor. We need it to describe the relation beetwen the parton density and
the saturation momentum for this case since we cannot trust the
geometrical estimates of the area of interaction in this case, as
discussed in \cite{KLNDA}.

\begin{figure}[h] 
\begin{tabular}{c c} 
\epsfig{file=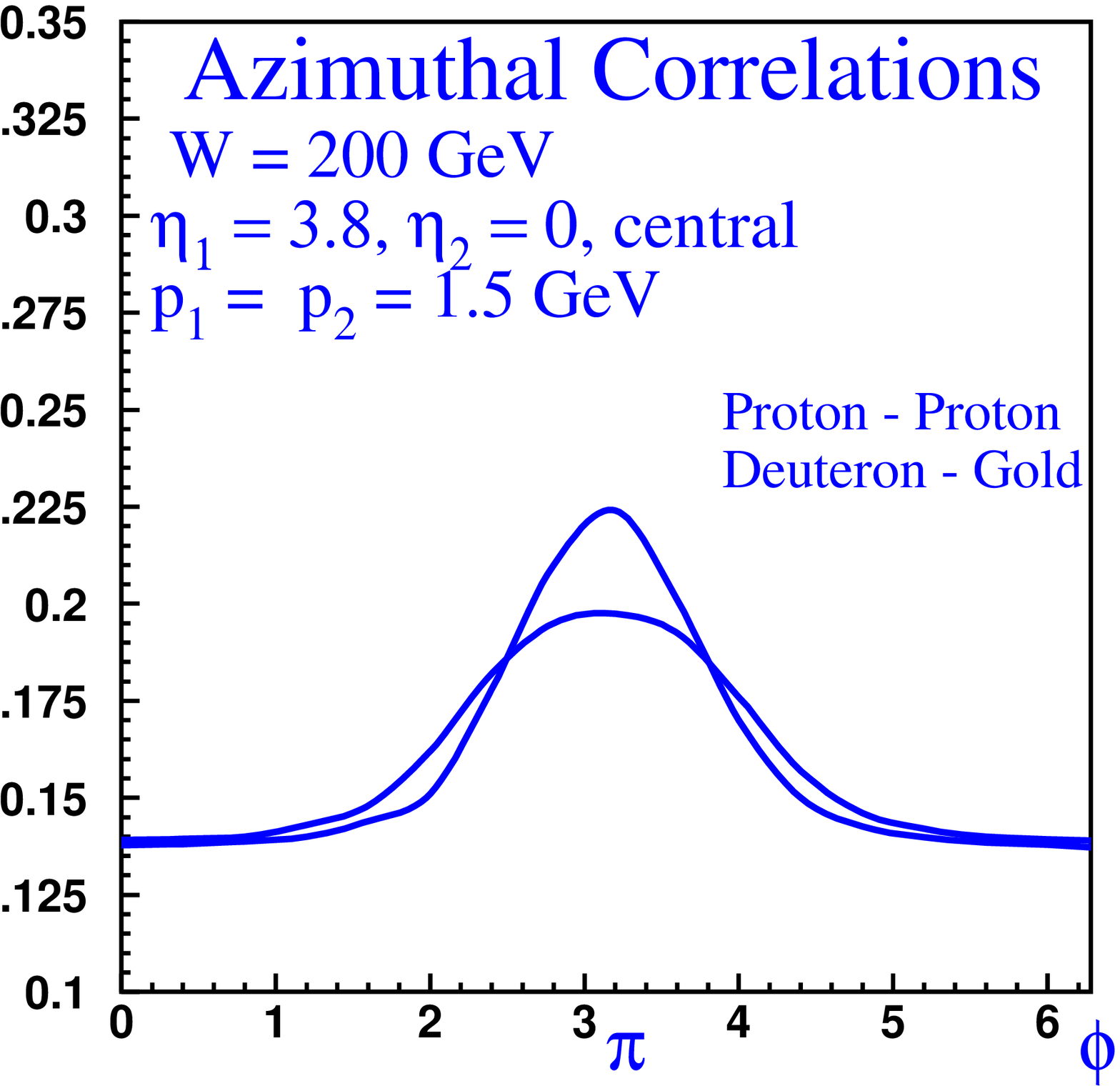,width=80mm} &
\epsfig{file=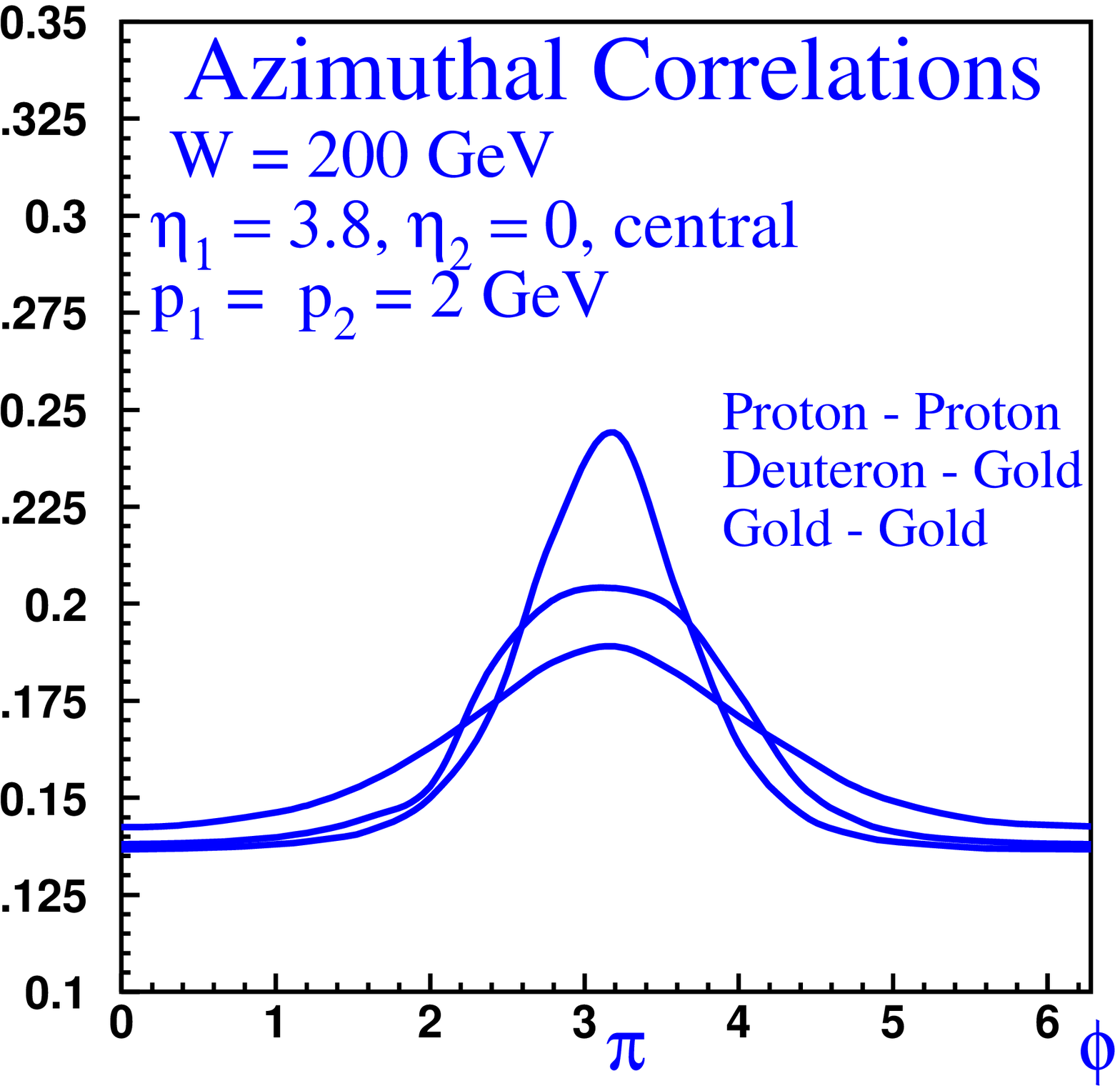,width=80mm} \\ 
\fig{corfor1}-a &
\fig{corfor1}-b \\ 
\end{tabular} 
\caption{\fig{corfor1}-a:  The azimuthal
angle correlations of produced jets at forward direction with rapidity
$\eta_2 =0$ and with $\,p_2\,\,=\,\,1.5\,GeV$ when the trigger jet is at
$\eta_1 = 3.8$ and with transverse momentum $p_1 = 1.5\,GeV$. Only
production from one parton shower is taken into account (see
\fig{cor0yfor}).  \fig{corfor1}-b: The azimuthal angle correlations of
produced jets at forward direction with rapidity $\eta_2 =0$ and with
$\,p_2\,\,=\,\,2\,GeV$ when the trigger jet is at $\eta_1 = 3.8$ and with
transverse momentum $p_1 = 2\,GeV$. Only production from one parton shower
is taken into account (see \fig{cor0yfor}).  } 
\label{corfor1}
\end{figure}

It is important to note that the uncorrelated production can be
subtracted from the data experimentally since the inclusive cross section
has been measured. Therefore, the correlation can be attributed to the
diagrams of \fig{cor0yfor}. In \fig{corfor1}-a and \fig{corfor1}-b we
illustrate the azimuthal angle distribution for two different kinematic
regions.

\begin{figure} 
\begin{center} 
\epsfysize=11cm 
\leavevmode 
\hbox{\epsffile{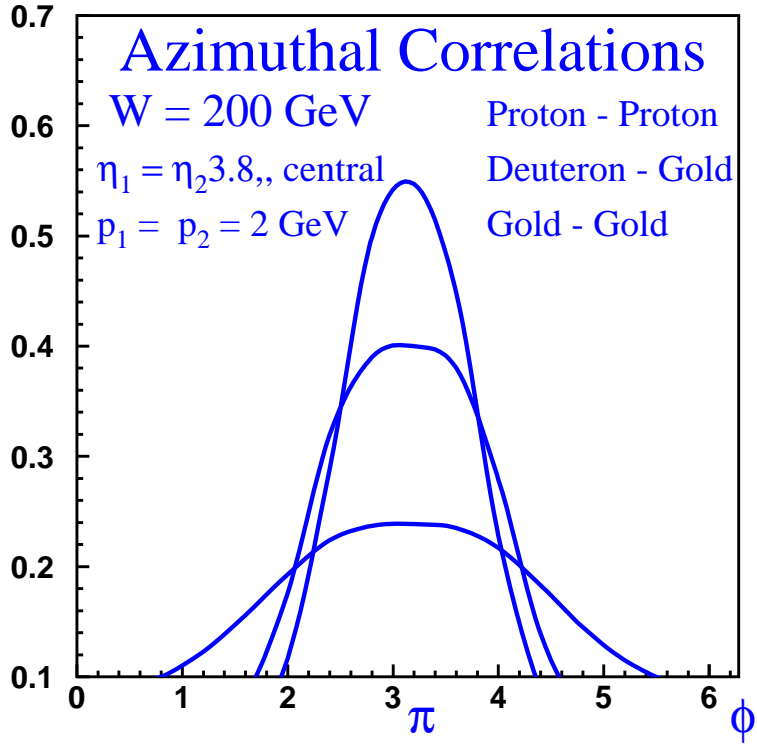}} 
\end{center} 
\caption{The azimuthal
angle correlations of produced jets at forward direction with rapidity
$\eta_2 =0$ and with $\,p_1\,\,=\,p_2\,\,=\,\,2\,GeV$ when the trigger jet
is at $\eta_1 = \eta_2\,=\,3.8$. Only production from one parton shower is
taken into account (see \fig{cor0yfor}).  }
\label{corfor3}
\end{figure}

These figures (\fig{corfor1}-a and \fig{corfor1}-b ) should be compared
with \fig{corfor3}, which gives the azimuthal correlations when both
triggers have the same rapidity ($\eta_1 = \eta_2 = 3.8$) and originate
from the same parton shower.

In fact one can substract the independent production of two hadrons just
by measuring the inclusive cross section. The remaining emission in one
parton shower then depends both on the saturation momenta for colliding
hadron and/or nuclei and on the BFKL emission in the rapidity interval
$\eta_1 - \eta_2$. The prediction for this process is theoretically
reliable and could provide the information on the values of the saturation
momenta.

To summarize, we have found that hadron azimuthal correlations provide a
stringent tests of the parton saturation in the Color Glass Condensate.  
When both hadrons are detected at central rapidity, our results show that
$pp$ and $dA$ correlations are quite similar, with some broadening in the
latter case, whereas the correlations in $AA$ are suppressed, mainly due
to the large independent production background. A most interesting
possibility is provided by the kinematics in which one of the hadrons is
detected at forward rapidity and another at central rapidity. In this case
the effects of quantum evolution in the CGC on the correlation function
are strong, and enhance the influence of the saturation boundary. This
leads to a significant difference between $pp$ and $dA$ correlation
functions.


\begin{thebibliography}{99}

\bibitem{RHICres1} R Debbe [BRAHMS Collaboration], Talk at the Fall 2003 DNP 
Meeting, Tucson, Arizona, October 2003, and at Quark Matter 2004 Conference, Oakland, California, January 2004;\\
I.~Arsene et al [BRAHMS Collaboration],
arXiv:nucl-ex/0403005.
\bibitem{RHICres2} 
M. Liu [PHENIX Collaboration], Talk at Quark Matter 2004 Conference, Oakland, California, January 2004; arXiv:nucl-ex/0403047.
\bibitem{RHICres3}
P. Steinberg [PHOBOS Collaboration], Quark Matter 2004 Conference, Oakland, California, January 2004.
\bibitem{RHICres4}
L. Barnby [STAR Collaboration], Quark Matter 2004 Conference, Oakland, California, January 2004.



\bibitem{KLM}
D.~Kharzeev, E.~Levin and L.~McLerran,
Phys.\ Lett.\  {\bf B561} (2003) 93
[arXiv:hep-ph/0210332].

\bibitem{KKT} D. Kharzeev, Y. Kovchegov and K. 
Tuchin, Phys.\ Rev.\ {\bf D68} (2003) 094013. 

\bibitem{Alb} J. L. Albacete, N. Armesto, A. Kovner, C. Salgado and U. 
Wiedemann, hep-ph/0307179 

\bibitem{BKW} R. Baier, A. Kovner and U. Wiedemann, 
Phys.Rev.D68:054009,2003

\bibitem{GLR}
L. V. Gribov, E. M. Levin and M. G. Ryskin, Phys. Rep. {\bf 100} (1983)
1.

\bibitem{hdQCD}
A.H. Mueller and J. Qiu, Nucl.Phys. {\bf B 268} (1986) 427;\\
J.-P. Blaizot and A.H. Mueller, Nucl. Phys. {\bf B 289} (1987) 847.

\bibitem{MV} L. McLerran and R. Venugopalan, Phys. Rev. {\bf D 49}
(1994) 2233; 3352; {\bf D 50} (1994) 2225.


\bibitem{CGC3}
E. Iancu, A. Leonidov and L. McLerran, Nucl.Phys.A692:583-645,2001; E. 
Iancu, E. Ferreiro, A. Leonidov and L. McLerran,
Nucl.Phys.A703:489-538,2002. 

\bibitem{IF}
E. Laenen and E. Levin, Ann. Rev. Nuc. Part. Sci. 44 (1994) 199;\,\,
Yu. V. Kovchegov and D. Rischke, Phys. Rev. C56  (1997)  1084;\,\,
M. Gyulassy and L. McLerran, Phys. Rev. C56 (1997) 2219;\,\,
Yu. V. Kovchegov and A. H. Mueller, Nucl. Phys. B529 (1998) 451\,\,
M. A. Braun, Eur. Phys. J. C16 (2000) 337, {\tt hep-ph/0010041},\,\,
{\tt hep-ph/0101070};\,\,
Yu. V. Kovchegov,  Phys. Rev. D64 (2000) 114016;\,\,
Yu. V. Kovchegov and K. Tuchin, Phys.\, Rev.\, {\bf D65} (2002) 074026
{\tt hep-ph/0111362}.

\bibitem{KN} D. Kharzeev and M. Nardi, Phys. Lett. {\bf B507} (2001) 121.
 
\bibitem{KL} D. Kharzeev and E. Levin, Phys. Lett. {\bf B523} (2001) 79;
nucl-th/0108006.
\bibitem{KLN}
D.~Kharzeev, E.~Levin and M.~Nardi,
{\it ``The onset of classical QCD dynamics in relativistic heavy ion  collisions,''}
{\tt hep-ph/0111315}.

\bibitem{KLNDA}

D.~Kharzeev, E.~Levin and M.~Nardi,
Nucl.\ Phys.\ A {\bf 730}, 448 (2004) and Erratum in arXiv:hep-ph/0212316.

\bibitem{RHICAZ} J. Adams, Phys. Rev. Lett. 91 (2003) 072304; C. 
Adler, Phys. Rev. 
Lett. 90 (2003) 082302


\bibitem{KT}
Y.~V.~Kovchegov and K.~L.~Tuchin,
Nucl.\ Phys.\ A {\bf 708}, 413 (2002)
[arXiv:hep-ph/0203213];  Nucl.\ Phys.\ A {\bf 717}, 249 (2003)
[arXiv:nucl-th/0207037].





\bibitem{AGK}
V.~A.~Abramovsky, V.~N.~Gribov and O.~V.~Kancheli,
Yad.\ Fiz.\  {\bf 18} (1973) 595
[Sov.\ J.\ Nucl.\ Phys.\  {\bf 18} (1974) 308].


\bibitem{MV1}
Yu.V. Kovchegov, Phys. Rev. {\bf D 54} (1996) 5463; 
J. Jalilian-Marian, A. Kovner, L. McLerran, H. Weigert, Phys.Rev. {\bf
D55} (1997) 5414;\,\,
E. Iancu and L. McLerran, Phys.Lett. {\bf B510} (2001) 145;\,\,
A. Krasnitz and R. Venugopalan, Phys. Rev. Lett.{\bf  84}  (2000)
4309;\,\,
E. Levin and K. Tuchin, Nucl.\,Phys. {\bf B573 }(2000) 833;
{\bf A693} (2001) 787; {\bf A691} (2001) 779;\,\,
A.H. Mueller,{\it ``Parton saturation: An overview,''} {\tt 
hep-ph/0111244};
E.~Iancu, A.~Leonidov and L.~D.~McLerran,
Nucl.\ Phys.\ A692 (2001) 583, {\tt hep-ph/0011241};
E. Iancu, K. Itakura and L. McLerran, Nucl. Phys. A708 (2002) 327, {\tt
hep-ph/0203137}.

\bibitem{MUKO}
Yu. V. Kovchegov and A. H. Mueller, Nucl. Phys. {\bf  B529} (1998) 451\,\,


\bibitem{SZC}
~A.~Szczurek,
{\it ``From unintegrated gluon distributions to particle production in hadronic 
collisions 
at high energies,''}
arXiv:hep-ph/0309146;\,\,Acta Phys.\ Polon.\ B {\bf 34} (2003) 3191
[arXiv:hep-ph/0304129].

\bibitem{STAR}
STAR collaboration: J. Adam et al., Phys. Rev. Lett. {\bf 91} (2003) 
072304.
\bibitem{GW} K. Golec-Biernat and M. W{\"u}sthof, Phys. Rev. {\bf D59}
(1999) 014017;
Phys. Rev. {\bf D60} (1999) 114023;
A. Stasto, K. Golec-Biernat and J. Kwiecinski, Phys. Rev. Lett. {\bf 86}
(2001) 596.\
\bibitem{KKP}
B.~A.~Kniehl, G.~Kramer and B.~Potter,
Nucl.\ Phys.\  {\bf B582} (2000) 514
[arXiv:hep-ph/0010289].

\bibitem{Les} A. Ogawa  (for the STAR Collaboration), AIP Conf. Proc. 675, 407 (2003).

\bibitem{MuellerN} A.H. Mueller and H. Navelet, Nucl. Phys. {\bf B 282} (1987) 727.


\bibitem{BFKL}
E.A. Kuraev,  L.N. Lipatov and V.S. Fadin, {\it Sov. Phys. JETP} {\bf 45},
 199       (1977); \\
 \,\,Ya.Ya. Balitskii and L.V. Lipatov, {\it Sov.
J.
Nucl. Phys.} {\bf 28}, 822  (1978);\,\,\\
L.N. Lipatov,      {\it  Sov. Phys.
JETP}
{\bf 63}, 904 (1986).



\end{thebibliography}
\end{document}